\documentclass[12pt]{article}
\usepackage{graphicx}
\begin{document} 

\title{Sornette-Ide model for markets: Trader expectations as 
imaginary part}

\author{Christian Schulze\\
Institute for Theoretical Physics, Cologne University\\D-50923 K\"oln, Euroland}

\maketitle
\centerline{e-mail: ab127@uni-koeln.de}

\bigskip

Abstract: A nonlinear differential equation of Sornette-Ide
type with noise, for a complex variable, yields endogenous
crashes, preceded by roughly log-periodic oscillations in the 
real part, and a strong increase in the imaginary part. The 
latter is interpreted as the trader expectation. 

 
\bigskip
Numerous microscopic models \cite{lls} for price fluctuations 
on stock or currency markets have been invented within the last
decade \cite{takayasu,rev} in the physics literature. An alternative are more 
phenomenological differential equations for the price itself
\cite{bouchaud,soride,chang,proykova,rosenow}, with some noise as
in Langevin equations. The
present note follows Sornette and Ide \cite{soride} but uses
a complex instead of a real variable. We interpret the real
part of this complex variable $z$ as the price (more precisely,
it is proportional to the logarithm of the price in units of 
the fundamental price), and the imaginary part can be the trader
expectation. Mathematics relates the changes in the real and
imaginary parts. We will try to see crashes arising from the 
intrinsic market forces, with (log-periodic \cite{sorjoh})
precursors in the real part and strong increases in the 
imaginary part. 

The expectations of the traders in the combination of price and expectation can 
be defined as a two-dimensional point in the market ``phase space'', which is
denoted by the complex number $z$ with Re$(z)$ = price and Im$(z)$ = 
expectation.

Thus, the differential equation for $z$ as a function of time $t$
is:
$$ d^2z/dt^2 = a\cdot e^z + b\cdot (dz/dt)^3 - c\cdot z^5$$
with $z$ at every time step changed by a small fraction 
$r\epsilon$ with $\epsilon \ll 1$ and a random number $-1<r<1$.
As before \cite{soride} the $c$-term incorporates fundamentalist
trading behaviour (buy if price is low), the $b$-term gives
herding (buy if others buy), and $r \epsilon$ the multitude of
new informations influencing the markets. Our new $e^z$ term
facilitates periodicities through the imaginary part of $z$
which part is thus tentatively identified with trader 
expectations. The exponents 3 and 5 in the above equations have
worked in previous simulations \cite{proykova} of a real $z$ 
at $a=0$. 

\begin{figure}[hbt]
\begin{center}
\includegraphics[angle=-90,scale=0.5]{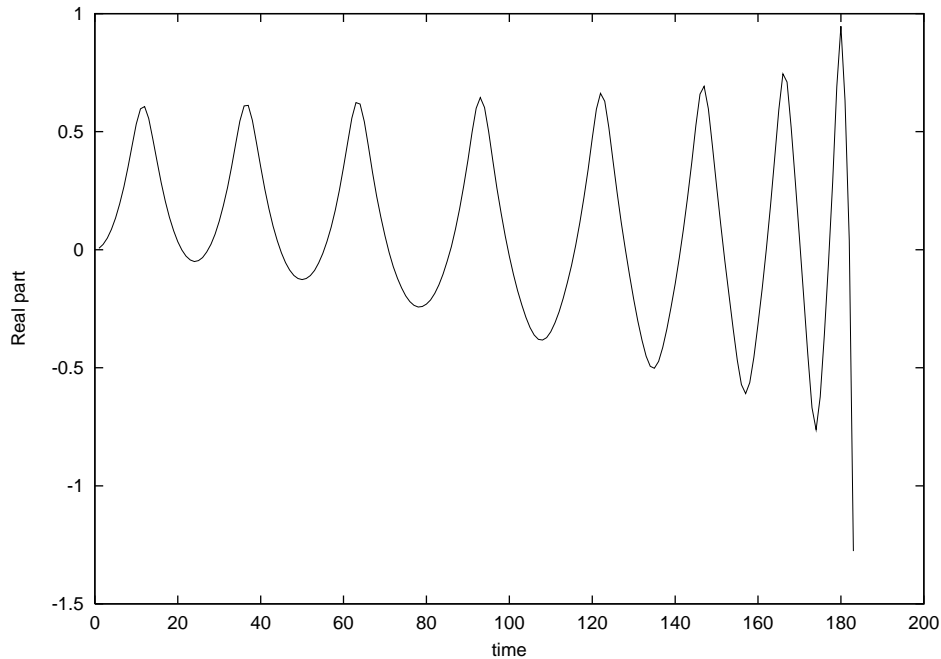}
\end{center}
\caption{
Real part of $z$ versus time up to the crash.
}
\end{figure}

\begin{figure}[hbt]
\begin{center}
\includegraphics[angle=-90,scale=0.5]{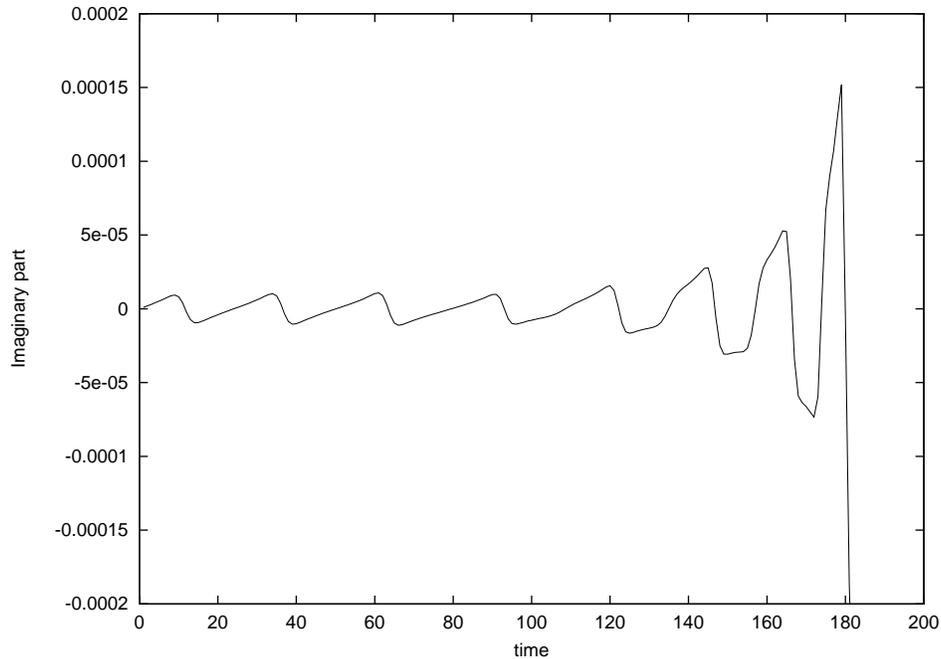}
\end{center}
\caption{
Same simulation as in Fig.1, but imaginary part of $z$ is shown instead.
}
\end{figure}

Figures 1 and 2 show one of our many results, for $\epsilon =
0.001, \, a = 0.01, \, b=c=1$, starting with $z = 0, \, dz/dt =
(1+10^{-3}i)10^{-3})$. We see in the real part a crash preceded
by oscillations of increasing amplitude and decreasing period, 
as in earlier work
\cite{soride,proykova}. The imaginary part in contrast becomes large
with changing signs only shortly before the crash, as it happens 
with trader panic and euphoria on real markets.

We thank Deutsche Forschungsgemeinschaft for support and D. 
Stauffer for help.

\end{document}